\documentstyle[12pt,epsf]{article}
\newcommand{\be}{\begin{equation}}
\newcommand{\ee}{\end{equation}}
\newcommand{\bea}{\begin{eqnarray}}
\newcommand{\eea}{\end{eqnarray}}
\newcommand{\bm}[1]{\mbox{\boldmath${#1}$}}
\newcommand{\ket}[1]{\mbox{$\left|\left.{#1}\right.\right \rangle$}}
\newcommand{\bra}[1]{\mbox{$\left.\left\langle{#1}\right.\right |$}}
\newcommand{\ba}[1]{\mbox{$\overline{#1}$}}
\newcommand{\inte}{\!\int\!\!d^3\!r\:}

\begin{document}
\title{RPA Description of the 
Electric Polarizability of the Nucleon
\footnotetext{$^\dagger$ Supported by GSI and BMBF}
}
\author{  J. Geiss, S. Hardt, H. Lenske and 
U. Mosel
\\
Institut f\"ur Theoretische Physik, Universit\"at Giessen$^\dagger$\\
D--35392 Giessen, Germany}
\maketitle
\date{}

\begin{abstract}
Excited states of the nucleon are described as RPA configurations on a
mean--field ground state taken from the MIT bag model. A residual
interaction of a structure as in the Nambu--Jona--Lasinio model is used.
The particle-hole states are coupled to good total angular momentum and
isospin.
Valence excitations of particle-hole type and quark--antiquark
($q\overline{q}$) states from the Dirac--sea are included.
The dependence of the baryon spectrum and
multipole response functions on the coupling constant $G$ is studied. At
critical values of $G$ the 3q--ground state becomes degenerate with
strongly collective $q\overline{q}$ modes. The model is used to
calculate the electric polarizability of the neutron $\alpha_N$. Without
residual interaction  $\alpha_N=7\cdot 10^{-4}fm^3$ is found. With
residual interaction the value increases to $\alpha_N=(7-11)\cdot
10^{-4}fm^3$. The measured value of $\alpha_N$ is reproduced within
experimental error bars.
\end{abstract}
\clearpage
\setcounter{page}{1}

\section{Introduction}
The electromagnetic polarizabilities are fundamental properties of
many--body systems. They are a direct measure for the response of a
bound system to an external perturbation and therefore provide direct
information on the internal structure and dynamical laws of the system.
{}From recent Compton scattering experiments \cite{1,2,3} the
polarizability of the neutron $\alpha_N=(12\pm 1.5 \pm 2)\cdot
10^{-4}fm^3$ \cite{2} was extracted. Initiated by these measurements 
considerable effort has been spent over the last years on theoretical
investigations of the polarizability. Since solutions of lattice QCD for
finite
systems are at a very early stage the theoretical calculations have to
rely on
effective hadronic models. Reponse functions are a sensitive measure for
the dynamical model used to approximate the fundamental QCD
interactions. Since the electromagnetic polarizability is determined by
processes of low energy--momentum transfer such investigations are
suitable extensions from investigations of the ground state to the
baryon spectrum at a couple of hundred MeV of excitation energy. Studies
of the polarizability therefore provide a possibility to test the
physical content of effective quark models of low energy QCD.
In the past calculations have been performed in  the Skyrme model
\cite{4,5,6}, the MIT-Bag model \cite{7,8} and its chiral extensions
\cite{9,10}, chiral soliton models \cite{11,12,13}, chiral perturbation
theory \cite{14} and non-relativistic constituent quark models
\cite{15}.  With quark degrees of freedom only the polarizability is
found to range between $\alpha_N\sim 8\cdot 10^{-4}fm^3$ in the MIT-Bag
model to $\alpha_N\sim 30\cdot 10^{-4}fm^3$ in the non--relativistic
quark models where a quark core root--mean--square (rms) radius of about
$<r^2>_{core}^{1/2}\sim 0.7fm$ was used. In the chiral quark models
where also mesonic degrees of freedom are taken into account the
electric polarizability is found to be dominated by the response of the
pion cloud to the electric field. In \cite{9} a value of $\alpha_N\sim
(8-10)\cdot 10^{-4}fm^3$ is obtained for $<r^2>_{core}^{1/2}\sim 0.6fm$
and the quark core contributes only about $10\%$ to $\alpha_N$.

At present the theoretical development is at a rather early stage and
the calculations have to be considered as exploratory. Very often, wave
functions are described in a rather restricted model space containing
only a few configurations and interactions are treated perturbatively.
The question arises to what extent a more complete description of
residual interactions will affect the results.  In a strongly
interacting system like a hadronic state in the low energy--momentum
regime dynamical correlations should be taken into account. In a
many--body system they can give rise  to qualitatively
new phenomena which are characterized by phase coherence of many
components. Wellkown examples are the giant resonances in nuclei which
are observed as strongly collective transitions in response functions.
Very likely, coherent motion and collectivity is also important in
hadron physics as indicated, for example, by results of the 
Nambu--Jona--Lasinio (NJL) model
\cite{16,klev}. In NJL model mesons are obtained as coherent
$q\overline{q}$ excitations of the vacuum \cite{klvw,bernard}
and the
nucleon was described as a solitonic solution of the field equations
\cite{Sol1,Sol2}.

In this paper a first attempt is made to investigate dynamical
correlations in a baryon of finite size. In the first place we are
interested in global properties of the nucleon close to the
ground state rather than in details of baryon spectroscopy. In this
energy--momentum region confinement is the prevailing mechanism. Because
at present a theory of confinement is still missing we  
treat confinement in a phenomenological way by using an effective
field theory. As
discussed in section 2.1 the model follows closely the NJL approach and
includes quark fields only. Confinement is enforced by an auxiliary
scalar potential chosen as in the MIT bag model \cite{mitbag}. The model
space
includes up ($u$) and down ($d$) quarks and the full range of color
states, i.e. we use the MIT-Bag model \cite{mitbag} with $SU_f(2)$
flavor and $SU_c(3)$ color symmetry. The nucleon ground state is chosen
as a 3--quark ($3q$) configuration on top of the filled Dirac sea. The
valence quarks are put into the lowest positive energy eigenstate of the
bag Hamiltonian. The ground state wave function is fully
anti--symmetrized and color--neutral and carries good total angular
momentum and isospin. The bag parameters are chosen such that ground
state properties as for example the root--mean--square ($rms$) radius of
the nucleon are reproduced.

In section 2.2 excited states of the nucleon and response functions are
considered. We are interested only in the changes of wave functions
relative to the ground state rather than in a calculation in absolute
terms. This allows to use the particle--hole picture which is well
established in many--body theory \cite{FW,Neg} and simplifies the
theoretical and numerical effort considerably. The excited states are
expanded into one particle--one hole ($1p1h$) type excitations of the
valence shell and $q\overline{q}$ excitations from the Dirac sea into
positive energy states. Both types of excitations are coupled to good
total internal angular momentum and isospin and, by construction, carry
boson
quantum numbers. The $q\overline{q}$ contributions
are of genuine field theoretical
character and are absent in a non--relativistic description as e.g. in
ref.\cite{nonr}.
 
Dynamical correlations of ph-states are investigated by
introducing an effective residual quark-quark interaction of the
NJL-type \cite{16}. The interaction is invariant with respect to Lorentz
transformations, rotations in color and flavor space and with
respect to chiral transformations. Thus, the
relevant symmetries of the QCD are accounted for. Certain matrix
elements of the residual interaction are found to be of the order of the
unpertubed excitation energies. Thus, a pertubative treatment as chosen
in other approaches \cite{17} is inappropriate. Using relativistic RPA
theory  residual interactions are included to all orders by exact
diagonalization within the model space. Moreover, the RPA description
also accounts for ground state correlations \cite{Rowe,FW,Neg,LW} which
are
known to be especially important for a reliable description of
collectivity.

Relativistic RPA theory is by itself a rather new field. Applications to
finite systems exist only for few cases in nuclear structure theory
\cite{drpa1,drpa2}. In section 2.3 we present a brief summary of the Dirac
RPA formalism and the application to baryonic excitations.

Our main interest is to study the response of baryons to probes with low
energy--momentum transfer. From the experience with nuclear response
functions it can be expected that correlations and possibly other
many--body effects will become observable. Of particular interest are
collective states which may lead to coherent excitations with mesonic
quantum numbers. This allows to study the evolution of mesonic modes as
dynamically correlated $q\overline{q}$-excitations in a baryonic system.

In section 3 numerical calculations and the application of the approach
to the polarizability of the nucleon are presented. In section 3.1 the
dependence of the energy spectrum on the model used for the ground state
and the residual interaction is discussed. The electrostatic
polarizability of the nucleon is investigated  in section 3.2. The
polarizability is determined from the inverse energy weighted RPA sum
rule. Vacuum polarization requires a special treatment. For a comparison
to the measured polarizability the physical value of $\alpha_N$ is
obtained by a RPA calculation including excitations of the valence
shells and the Dirac sea from which the polarizability of the bare
vacuum is subtracted. The latter is obtained by an independent RPA
calculation without valence quark states. In this paper we are mainly
interested in the electric polarizability of the neutron. We find that
$\alpha_N$ is increasing when switching on the residual interaction. We
obtain for a quark core root--mean--square radius (rms) of about $0.65
fm$ $\alpha_N\approx 7\cdot 10^{-4}fm^3$ for the pure MIT-Bag in
agreement with \cite{9} and $\alpha_N\approx 10.5\cdot 10^{-4}fm^3$ when
including the residual interaction. The paper closes in section 4 with a
summary and concluding remarks.

\section{Effective Theory of Hadron Structure}
\subsection{Nucleon Ground State and Effective Hamiltonian}
In this section an effective quark field theory of a dynamical structure
similar to the Nambu--Jona--Lasinio (NJL) model \cite{16} is discussed.
The NJL model can be considered as a model for the quark sector of QCD.
Gluonic contributions are integrated out. They are treated schematically
in terms of a chirally symmetric two--body contact interaction. Of special
interest for hadron physics is that chiral symmetry is spontaneously
broken in the NJL model. For coupling constants beyond a critical value
a sudden transition from massless quarks to heavy  constituent quarks is
observed. Thus, the NJL model explains the chiral phase
transition dynamically. It is also well suited for model studies of
dynamical correlations \cite{klvw,klev}, medium effects and solitonic
states
\cite{Sol1,Sol2}. A review of the NJL model and applications is found in
ref.\cite{klev}.  For our purpose we use a model Lagrangian ${\cal L}$
\bea\label{lagra}
{\cal L}&=&{\cal L}_{MF}+{\cal L}_{int}\\
{\cal
L}_{MF}&=&\overline{\Psi}\left(i\gamma_{\mu}\partial^{\mu}-m_q-U_s\right)\
Psi\\
{\cal L}_{int}&=&{\cal L}^c_{int}+\ba{\Psi}U_s\Psi\\
{\cal L}^{c}_{int}&=&G\sum_{i=1}^8 \left( \ba{\Psi}\gamma_{\mu}
{\lambda^i\over 2}\Psi \right)^2.
\eea
given by a mean--field part ${\cal L}$$_{MF}$
and a residual interaction including a two--body
color--color interaction ${\cal L}$$^c_{int}$.
An auxiliary scalar potential $U_s$ has been introduced 
in order to obtain confinement. In practice confinement is described in
the
MIT bag model which means to set $U_s=const.$ inside the bag and to impose
appropriate confining boundary conditions on the bag surface. The quark
current mass
is denoted by $m_q$.

The interactions $\cal{ L}$$^c_{int}$ is given by a scalar coupling of
color currents.
The color structure is determined by the generators $\lambda_i, i=1\cdots
8$
of $SU_c(3)$ \cite{Itz}. In a color neutral state the matrix elements of
such an interaction
are completely determined by the exchange contributions. For a contact
interaction
as in ${\cal L}$$^c_{int}$ an equivalent direct interaction is obtained
after a
Fierz transformation \cite{Itz} which for $SU_f(2)$ is given by
\begin{equation}
{\cal L}^c_{int}=\sum_{k=S,V\cdots}{\left(
a_k (\overline{\Psi}\Gamma_{k}\Psi)^2+
b_k (\overline{\Psi}{\bm \tau}\Gamma_{k}\Psi)^2\right)}  
\end{equation}
where $\Gamma_k=1,\gamma_\mu \cdots$ is a scalar, vector etc. Dirac
operator
and ${\bm \tau}$ is a Pauli isospin operator.
The coefficients $a_k$ and $b_k$ are related to the coupling constant $G$
by
the elements of the Fierz transformation matrix.

The stationary solutions to the mean--field Lagrangian ${\cal L}$$_{MF}$
are given by the eigenstates of the Hamiltonian $H_{MF}=H_{bag}$
\bea
\left[ \bm{\alpha p}+\beta M(\bm{r})- E_{n,\kappa}  \right]
\Psi_{n,\kappa,j}(\bm{r})=0
\eea
where confinement is described by the effective mass operator
\bea\label{mass}
M(\bm{r})=m_q+U_s(\bm{r})=\left\{
\begin{array}{rl} M_0; & r<R \\ \infty; & r>R \end{array}
\right. 
\eea
and the constant $M_0$ can be set to zero.
$\kappa,j$ are the eigenvalues of the operators  $K=\beta(\bm{\Sigma \cdot
\ell} +1)$ and the total angular momentum $\bm{j}=\bm{\ell}+\bm{s}$. The
quark
wave functions can be written in the form
\bea
\Psi_{n,\kappa,j}=\left(\begin{array}{r}
g_{n,\kappa}(r) \mbox{$\cal Y$}_{j l}^m \\
-if_{n,\kappa}(r) \mbox{$\cal Y$}_{j l^{\prime}}^m
\end{array}\right) e^{-iE_{n,\kappa}t}.
\eea
Quantization is obtained by the condition that the quark currents vanish
at the surface of the bag \cite{mitbag}. 
In $SU_f(2)$ the lowest positive energy level has quantum numbers
$j^{\pi}={1 \over 2}^{+}$, $t={1 \over 2}$ and is therefore fourfold
degenerate. The ground state of a baryon is built of three quarks in the
valence shell
which are coupled to good total angular momentum and parity $J^\pi$ and
isospin $T$.
In addition, the state must be a $SU_c(3)$ color singlet. In second
quantization
the wave function of a neutron with spin projection $m_j=+{1 \over 2}$
is given by \cite{mosel}
\bea
\ket{n\uparrow}={1\over \sqrt{18}} \sum_{\alpha\beta\gamma=1}^4
T_{\alpha\beta\gamma} {a_{\alpha}^1}^{\dagger} {a_{\beta}^2}^{\dagger}
{a_{\gamma}^3}^{\dagger}\ket{0}.
\label{neutron1}
\eea
The bare vacuum is denoted by $\ket{0}$
and the states in color space are labeled by $1,2,3$.
The summations over $\alpha,\beta,\gamma$ account for the
two flavors and the spin projections where indices have been assigned
\bea\label{index}
\ket{u\uparrow}\mapsto 1\qquad \ket{u\downarrow}\mapsto 2\qquad
\ket{d\uparrow}\mapsto 3\qquad \ket{d\downarrow}\mapsto 4\quad .
\eea
The coefficients $T_{\alpha\beta\gamma}$ are given by
\bea
T_{\alpha\beta\gamma}=\left\{
\begin{array}{rl} -1;& (\alpha,\beta,\gamma)=(1\,3\,4) \quad + \quad
\mbox{permutations} \\
2;& (\alpha,\beta,\gamma)=(2\,3\,3)\quad + \quad
\mbox{permutations} \\
0; & \mbox{otherwise}.
\end{array}
\right.
\eea
{}From the energy--momentum tensor $T^{\mu\nu}$ the ground state energy
is obtained as
\be\label{egs}
E_{gs}=\langle n|T^{00}|n \rangle=\sum_{k}{E_{k} g(k)}
+\langle n|\ba{\Psi}U_s\Psi|n\rangle+\langle n|{\cal L}_{int}|n \rangle
\ee
where the mean--field energies $E_k$ are weighted by the ground state
occupation
probabilities
\bea
\bra{n\uparrow} a_k^{\dagger} a_{k'} \ket {n\uparrow}
=\delta_{k,k'} \cdot g(k)=\delta_{k,k'}\cdot
\left\{ \begin{array}{rl} 1, & E_k<0 \\ 0, & E_k > E_{val} \\
{1 \over 18}g_v(k),& E_k=E_{val}. \end{array} \right.
\label{occupation}
\eea
and
\bea
g_v(1)=2 \qquad g_v(2)=4 \qquad g_v(3)=10 \qquad g_v(4)=2
\eea
where the same labeling as in Eq.(\ref{index}) is used.
 
The use of the phenomenological potential $U_s$ clearly describes
confinement in a non--self--consistent way. Translational and chiral
symmetry is broken by $U_s$ and from Hartree-Fock theory it is known
that  this may lead to admixtures of spurious states into the spectrum.
This occurs in channels with the same quantum numbers as the
broken symmetry, for example the spurious centre of mass motion in the 
isoscalar $J^\pi=1^-$ channal.   By
imposing the condition that the residual interaction does not contribute
to the ground state the depth of the confining potential or equivalently
the constant $M_0$, Eq.(\ref{mass}), can be fixed
\be\label{depth}
U_s=-2a_s\langle n|\overline{\Psi}\Psi|n \rangle
\ee
where $a_s$ is the scalar coupling constant after the Fierz
transformation. By this relation
self--consistency between the auxiliary potential $U_s$ and the two--body
interaction
is restored at least at an average.

After the ground state has been obtained we next consider excited states
of the nucleon. For that purpose it is convenient to extract an
effective Hamiltonian from the energy--momentum tensor.
We arrive at a form which can be written as
\bea\label{heff}
H=\sum_k E_k a_k^{\dagger} a_k+{1 \over 4}\sum_{pqrs} V_{prqs}
:a_p^{\dagger}
a_r^{\dagger} a_s a_q:,
\eea
where
$E_k$ are the MIT-bag eigenenergies and the normal ordering is
taken with respect to the nucleon groundstate.
The matrix elements of the residual interaction are defined by
\bea\label{vers}
V_{prqs}=G\inte \left( \ba{\varphi}_p
\gamma_{\mu}\frac{\lambda_{\alpha}}{2}
\varphi_q \right)\left( \ba{\varphi}_r
\gamma^{\mu}\frac{\lambda_{\alpha}}{2}
\varphi_s \right).
\eea
The summations include contributions from positive and negative energy
states.
An energy cut--off $\Lambda$ in the ph-space has to be introduced. 
This corresponds to the fact, that the interaction of Eq.(\ref{vers})
would have to be
regularized in case the mean-field had been extracted self-consistently.
A cut--off is also necessary by numerical reasons
in order to keep the calculations manageable.

\subsection{Excited States and Response Functions}
An appropriate approach to small amplitude oscillations 
is to represent excited states of the nucleon by particle--hole
excitations with respect to
the ground state. This means to use the ground state as a new vacuum and
to assume that
the wave functions of excited states are mainly determined by a
redistribution of 
single particle
strength from occupied into initially empty states. In a relativistic
system this redistribution
process also includes transitions from the Dirac sea into positive energy
states. The ground
state might contain admixtures of many--body configurations beyond the
simple mean--field
state. In that case excited states also include components from the
annihilation
of such ground state correlations \cite{Rowe,LW}. 
In RPA theory this is taken into account by using state operators
\bea
Q_{\nu}^{\dagger}(J,\!M,T,\!M_T)\!=\!\sum_{ph}\left[
X_{ph}^{\nu}A^{\dagger}_{ph}(J,\!M,T,\!M_T)\!-\!
Y_{ph}^{\nu} A_{ph}(\overline{J},\!M,\overline{T},\!M_T)\right].
\label{rpa}
\eea
which include $1p1h$ excitation operators  
\bea\label{tfw}
A^{\dagger}_{ph}(J,M,T,M_T)=(a_p^{\dagger} \otimes a_h)^{J,T}_{T,M_t}
\eea
and annihilation operators
\bea\label{tbw}
A_{ph}(\overline{J},M,\overline{T},M_T)
=(-1)^{J+M+T+M_t}(a_h^{\dagger} \otimes a_p)^{J,-M}_{T,-M_t}
\quad .
\eea
The particle and hole operators $a_p^{\dagger}$
and $a_h$ are coupled to total angular momentum
$J$ and isospin $T$ with projections $M$ and $M_T$, respectively.
The structure of the RPA state operators and their dynamics will be
discussed in the next section. Here, we consider first their application
to excited states of the nucleon. Denoting the ground state by
$|j_0m_0\rangle$
excited states are given in general by a superposition of RPA multipole
operators
\be\label{jm}
|jm,tt_3\rangle=\sum_{J,T}{\left[Q^{\dagger}(J,T)\otimes|j_0,t_0
\rangle\right]_{jm,tt_3}}
\ee
Since the RPA operators carry bosonic quantum numbers it is seen that
excited states are actually expressed by a superposition of 
valence excitations and meson-like
modes. The wave functions of Eq.(\ref{jm}) lead to the so--called tensor
RPA \cite{18} which will not be discussed here. An important
simplification
is obtained if only multipole response functions and sum rules are
considered.
In that case one is only interested in the total strength for a transition
of a given multipolarity regardless of the spin states $(jm,tt_3)$ which
were
participating in the excitation process.
This amounts to average over the spin and isospin projections of the
initial state
 $(j_0m_0,t_0{t_3}_0)$
and to sum over final spin and isospin quantum numbers $(jm,tt_3)$. 
The spin averaging corresponds
to project out the bosonic RPA modes of the same tensorial rank
as the operator of the external probe.
In electric dipole transitions, for example,
only the $J^{\pi}=1^{-}$ components of the final states contribute.
Therefore, spin averaged response functions and
sum rules are described to a good approximation already in the
uncoupled representation
\be\label{uc}
|j_0m_0,t_0{t_3}_0;JM,TM_T\rangle=Q^{\dagger}(JM,TM_T)
|j_0m_0,t_0{t_3}_0\rangle
\quad .
\ee
and these wave functions will be used in the following.

\subsection{Dirac RPA Theory}
The RPA state operators, Eq.(\ref{rpa}), are of a particular structure
which goes beyond a pure mean--field description. The terms
containing the annihilation operators correspond to processes which
propagate
{\em backward} in time. Such processes only contribute if the ground state
includes
correlations of $2p2h$ or higher order.
The correlated RPA ground state $|0_c\rangle$ is defined by the
orthogonality
condition
\be\label{ortho}
Q_\nu |0_c\rangle=0
\ee
which must be fulfilled for all RPA states $\nu$.
As shown in ref.\cite{LW} the RPA and the mean--field ground states
are related by
\bea\label{rpags}
|0_c\rangle=e^S |0\rangle
\eea
where correlations are introduced by the operator
\be\label{corr}
S=\sum_{ph,p'h',kk_z}{C_{php'h'}(k,k_z)A^\dagger_{ph}(k,k_z)A^\dagger_{p'h
'}(k,-k_z)}
\ee
from which the many--body structure of $|0_c\rangle$ is apparent. 
For a shorthand notation the combined spin--isospin quantum 
numbers $(k,k_z)=(JM,TM_T)$ were introduced.
Because the correlation
matrix $C$ is defined in terms of the "backward" amplitudes $Y$ \cite{LW}
RPA can be
solved self--consistently by successive recalculation of the ground state.
However, here we follow
the common practice and determine the $Y$--amplitudes in first
approximation which
corresponds to the quasi--boson approximation (QBA). It means to retain in
Eq.(\ref{rpags}) the lowest order term in $S$ when matrix elements and
occupation probabilities involving the time--reversed state operators are
to be evaluated.

In relativistic RPA the ground state correlations also include
$q\overline{q}$ admixtures from virtual excitations of the vacuum. In
nuclei these contributions are suppressed to a large extent
\cite{drpa1,drpa2} because of the large mass gap. A qualitatively
different situation is encountered in quark models of hadrons.  Klimt et
al. \cite{klvw} have used the NJL model to calculate the meson states
below $1~GeV$ as collective $q\overline{q}$ excitations of the
vacuum. From the results of ref.\cite{klvw} it can be expected that
$q\overline{q}$
vacuum correlations are of importance also for baryons.
Physically, they describe the dressing of the quark core  by a virtual
meson cloud due to dynamical vacuum polarization.

The RPA state operators describe stationary eigenstates of the full
Hamiltonian, Eq.(\ref{heff}). They are determined by the
{\em equation of motion} \cite{Rowe,FW}
\be\label{eom}
\left[H,Q^\dagger_\nu\right]=\omega_\nu Q^\dagger_\nu
\ee
where $\omega_\nu =E_\nu -E_{gs}$ is the excitation energy.
Taking the commutator from the left with resepct to the $1p1h$ excitation
and annihilation operators, Eqs.(\ref{tfw}) and (\ref{tbw}), the RPA
eigenvalue 
equation
\bea
\left( \!\!\begin{array}{cc}
	A(k,k_z) & B(k,k_z)\\
	B^{*}(k,-k_z) & A^{*}(k,-k_z)
       \end{array}\!\!
\right)
\left(\!\! \begin{array}{c}
       X^{\nu} \\ Y^{\nu}
       \end{array} \!\!
\right) = \omega_{\nu}
\left( \!\!\!\begin{array}{cc}
	N & 0\\
	0  & \overline{N}
	       \end{array}\!\!\!
\right)
\left( \!\!\begin{array}{c}
       X^{\nu} \\ Y^{\nu}
       \end{array}\!\!
\right)
\label{rpagleichung}
\eea
is obtained from which $\omega_\nu$ and the state vectors
$(X_\nu,Y_\nu)$ are derived.  
The matrix elements are defined by ground state expectation values
of double commutators
\bea
A_{mi,nj}(k,k_z)=\bra{0} [ A_{mi}(k,k_z),
[H,A^{\dagger}_{nj}(k,k_z) ] ] \ket{0}
\nonumber
\eea
\bea
B_{mi,nj}(k,k_z)=-\bra{0} [ A_{mi}(k,k_z),
[H,A_{nj}(\overline{k},k_z) ] ] \ket{0}
\nonumber
\eea
and the metric is given by
\bea
N_{mi,nj}(k,k_z)=\bra{0} [ A_{mi}(k,k_z),
A^{\dagger}_{nj}(k,k_z) ] \ket{0}
\nonumber
\eea
\bea
\overline{N}_{mi,nj}(k,k_z)=\bra{0} [ A_{mi}^{\dagger}(\overline{k},k_z),
A_{nj}(\overline{k},k_z) ] \ket{0}.
\eea
$m,n$ ($i,j$) denote single quark states above (below) the Fermi-surface.

Evaluating the commutators the metric $N$ is found to be given by
the expectation values
\bea\label{occ}
N_{mi,nj}&=&
\bra{0}\delta_{i,j}\delta_{m,n}-\delta_{i,j}a^\dagger_n a_m -
\delta_{m,n}a_j a^\dagger_i\ket{0}\\
\overline{N}_{mi,nj}&=&
\bra{0}-\delta_{i,j}\delta_{m,n}+\delta_{i,j}a^\dagger_n a_m +
\delta_{m,n}a_j a^\dagger_i\ket{0}
\quad .
\eea
In an open shell system like the nucleon where the valence shell is only
partially occupied the metric is determined by the ground state
occupation probabilities $g(k)$, Eq.(\ref{occupation}),
\be
N_{mi,nj}=-\overline{N}_{mi,nj}=\delta_{i,j}\delta_{m,n}(1-g(i)-g(m))
\quad .
\ee
In a system with a fully occupied valence shell the contributions from the
normal ordered
operator parts vanish and the standard RPA metric is retrieved:
\bea
N_{mi,nj}=-\overline{N}_{mi,nj}=\delta_{i,j}\delta_{m,n}
\quad .
\eea

\section{Results and Discussion}

The model accounts for a number of important features of hadron
structure physics, at least in the quark sector, and it is therefore
meaningful to calculate observables and compare them to data. 
In view of the phenomenological description of
confinement, the schematic {\em ansatz} for the residual interaction and
the restricition to the quark sector we consider the results as
exploratory. However, finite size effects are taken into account and RPA
theory allows to study residual interactions in a relativistic invariant
formulation.

\subsection{Energy Spectrum}
The ground state of the nucleon was described in MIT bag model where a
bag radius $R_{bag}=0.9~fm$ was used which corresponds to a rms
radius  $\langle r^2\rangle^{1/2}_c=0.65~fm$ for the quark core. A
value less than the empirically known proton charge radius $(0.8~fm)$ is
used because the model clearly does not account for contributions from
the meson cloud surrounding the core.  The spectrum of $1p1h$ states
obtained in the bag model is known to give a too high level density at
low excitation energy \cite{17} and, even more severe, the lowest excited
states are of negative parity (see Fig.1). This is in clear contradiction
to
experiment where the positive parity $\Delta(1232)$ resonance is found
as the first excited state above the nucleon. In terms of the
transferred angular momentum, as introduced in section 2.2, this is a
$J^\pi=1^+$ "pionic" excitation on the nucleon ground state. Within the
bag model a more realistic level sequence could not be obtained. The
results simply reflect the properties of the spectrum of an infinite
well where the orbits group into shells of alternating parity. Spectra
obtained with confining potentials of different shapes, e.g. linearily
increasing with radius, led to similar results. As it might have been
expected a bag model accounts only in a very rough way for the dynamics
of quarks in a nucleon.

The question arises whether the residual interaction could improve on this
situation. In the RPA calculations excitations from the valence shell
and the Dirac sea with unperturbed energies up to $2200~MeV$ where used
which corresponds to configuration space of about $N=40$ unpertubed
$1p1h$ and $q\overline{q}$ states. The dependence of the RPA
eigenenergies on the coupling constant $G$ of the residual interaction,
Eq.(\ref{lagra}), is illustrated in Fig.1 for isoscalar
$J^{\pi}=0^{\pm}$ and isovector $J^{\pi}=1^{-}$ states. Since these
channels involve spin and isospin modes, respectively, they are free of
admixtures from spurious translational states. The special form of the
residual
interaction leads after the Fierz transformation to isoscalar ($T=0$)
and isovector ($T=1$) interactions of the same structure and interaction
strengths. Thus, for given $J^\pi$, isoscalar and isovector excitations
behave in the same way when $G$ is varied and it is sufficient to
consider only one of the two channels.

Over a wide range of interaction strengths the eigenenergies remain on
their unperturbed values and seem to be almost unaffected by
the residual interaction. An inspection of the RPA matrix shows that the
non--diagonal elements of the RPA matrix are rather large and may reach
values of the order of the lowest $ph$ energies. Thus, strong
cancellations must occur which compensate the contributions from the
diagonal matrix elements. For $G>6\cdot 10^{-6}MeV^{-2}$ the first
$J^{\pi}=0^{-}$ state is lowered and approaches the ground state
($\omega=0$) at $G_{crit.}=6.8\cdot 10^{-6}MeV^{-2}$. At even larger
values
of $G$ the energy becomes purely imaginary which is a wellknown property
of the RPA \cite{Rowe}. On the way down to zero excitation energy the
states
gains collectivity and an increasing number of configurations
contribute. When become degenerate with the ground state the ground
state correlations strongly increase which is seen from the fact that
the $Y$ and $X$ amplitudes are of the same magnitude. 
Close to
this value of $G$, however, the RPA ceases to be
valid and, as discussed in section 2.3, an improved description of the
RPA ground state would be necessary.
These transition depends to some extend on the
multipolarity. In Fig.1 also results for the $J^\pi=0^+$ channel are
shown. In that case $G_{crit.}=6.4\cdot 10^{-6}MeV^{-2}$ is found. In the
$J^\pi=1^-$ channel the cancellation effects extend to larger coupling
constants. As a result the states are less affected by residual
interactions in the region of $G$ values shown in Fig.1.

The results lead to the conclusion that a perturbative
treatment of residual interactions in a
relativistic system as e.g. in ref.\cite{17} might be unreliable. 
The RPA results show that an
exact diagonalization can lead to cancellations such that even if the
matrix elements are individually large the net effect is minor. 

Calculations with larger configuration spaces led to qualitatively the
same results. The dependence on $G$ roughly scales with the size of the
configuration space. The results indicate that the product of G and the
square of the cut--off energy $\Lambda$ is close to a constant. A
similar scaling law is found in standard NJL calculations with a three--
or four--momentum cut--off $\Lambda_p$ where observables depend only on 
$G \Lambda_p^2$.

\subsection{The Electromagnetic Polarizability}
The electric and magnetic polarizabilities $\alpha$ and $\beta$,
respectively, are a measure for the response of a system to a time
independent external electromagnetic field. Theoretically, they are
defined by the inverse energy weighted sum rule \cite{rs}
\bea
\alpha=2\cdot \sum_{\nu \ne 0} {\left| \bra{0} D_z \ket{\nu} \right|^2
\over E_{\nu}-E_0},
\label{pol1}
\eea
where the ground state and the RPA excited states are denoted by
$\ket{0}$ and $\ket{\nu}$, respectively. The transition operator is
obtained by minimal coupling of the quark fields to the electromagnetic
field. The electric polarizability is determined by the dipole operator
\bea\label{dz}
D_z&=&e\int \ba{\Psi}\gamma_0 A_0 \Psi d^3r\\
A_0&=&z=\sqrt{{4\pi\over 3}} r Y_1^0
\eea
For the nucleon the RPA matrix elements of $D_z$ are given by
\bea\label{dmat}
\bra{n} D \ket{\nu} =\sum_{ph} \bra{p} D \ket{h}
\left(X_{ph}^{\nu}+Y_{ph}^{\nu}\right)(g(h)-g(p))\quad .
\eea
The angular momentum and isospin coupled matrixelement of $D_z$
is denoted by $\bra{p} D \ket{h}$. The summation includes $1p1h$ valence
and
$q\overline{q}$ vacuum excitations.  $g$ are the occupation numbers,
Eq.(\ref{occupation}), averaged over the spin projections $M$. From the
transition matrix elements it is seen that  the structure of the RPA
wave functions is directly tested by the polarizability.

A special feature of relativistic systems is vacuum polarization. The
contributions corresponding to $q\overline{q}$ excitations of the bare
vacuum  have to be removed from Eq.(\ref{pol1})  before a quantity is
obtained which can be compared to data. The physical polarizability is
determined by subtracting the polarizability $\alpha_{vac}$ of the bare
vacuum from the polarizability $\alpha_{vac+3q}$ of the full system,
\bea
\alpha_N=\alpha_{vac+3q}-\alpha_{vac}\quad .
\eea

Both values $\alpha_{vac+3q}$ and $\alpha_{vac}$ are obtained by
RPA calculations on the bag groundstate.

In Fig.2  $\alpha_{vac+3q}$ and $\alpha_{vac}$ are
shown as a function of $G$. Both values are seen to increase with
the strength of the residual interaction. This behaviour reflects the
gain in collectivity with increasing interaction strength which is
typical for RPA response functions. A collective state is obtained
when a large number of basis states are phase coherent. A special
property of the RPA is the additional increase in transition strength
from the ground state correlations. In a collective state the $y$ and
$x$ amplitudes carry the same sign and Eq.(\ref{dmat}) shows that
matrix elements will considerably
be enhanced. The effect is especially pronounced for low--lying state
which are close to the ground state and thus are most strongly affected by
ground state correlations.

At larger $G$ collectivity becomes important in $\alpha_{vac+3q}$ as seen
from the increasing separation between the two curves in Fig.2.
Apparently, the increase in collectivity is stronger for the valence
components than
for the vacuum excitations.
The effect is seen more clearly in $\alpha_N$, displayed
in the lower part of Fig.2. Over the range of coupling constants
$\alpha_N$
increases by about 50\%. The gain in transition strength indicates the
gradual built--up of coherence in the RPA wave functions
and the valence
excitations participate more strongly than the $q\overline{q}$
configurations.
This result shows
that the electrostatic polarizability is a measure
for the properties of the quark core being less sensitive to vacuum
effects.
For $G>3\cdot 10^{-6}~MeV^{-2}$.
the theoretical results range within the error band of the measured
value $\alpha_N=(12\pm 1.5 \pm 2)\cdot 10^{-4}fm^3$ \cite{2}.
Thus the observed polarizability is reproduced reasonably well.

\section{Summary and Conclusions}
An effective field theoretical model was used to study dynamical
correlations in excited states of the nucleon. Starting from
a Lagrangian of a structure similar to the NJL model an
effective  Hamiltonian was derived. Confinement was
introduced by means of an phenomenological scalar potential. In
applications
the MIT bag model was used.
The ground state of the nucleon was described by color--singlet wave
function with good total angular momentum
and isospin. 
A two--body color--color interaction was used in relativistic RPA
calculations on a nucleon ground state.
In exploratory model calculations
the experimentally known electrostatic polarizibility of the nucleon
was studied. The empirical value was
reasonably well reproduced.
Similar to ref. \cite{9} we are led to the conclusion that the coupling of
the external photon
to mesonic modes which in our approach are represented
by collective RPA excitations gives important contributions to the
polarizability.
The model calculations indicate a special sensitivity of the
polarizability
to the quark core. The agreement
supports at least the physical concept of the model. However, in view of
the
remaining uncertainties it is
hardly possible to draw final conclusions on details of the dynamics and
other physical aspects.
The results for energy spectra and the
polarizability constrain the coupling constant to the region
$G=3\cdots6\cdot10^{-6} MeV^{-2}$.

\clearpage
\centerline{\bf FIGURE CAPTIONS}

\vspace{1cm}

\noindent
Fig. 1.

\noindent
Energy-spectra in the $J^{\pi}=0^{-},1^{-},0^{+}$ channels as
a function of the coupling constant $G$. In all cases only the first five
excited states are shown
\label{fig1}

\vspace{5mm}
\noindent
Fig. 2.

\noindent
Polarizability of the neutron as a function of the coupling strength
of the residual interaction. In the upper part RPA results for the
vacuum polarizability $\alpha_{vac}$ (dashed) and the polarizability
$\alpha_{vac+3q}$ including valence
states (full line ) are shown. In the lower panel the empirical (dashed)
and the theoretical polarizability  (full line)
of the neutron are displayed. Additionally the lower bound of the
empirical
value is shown (dotted line).
\label{fig2}

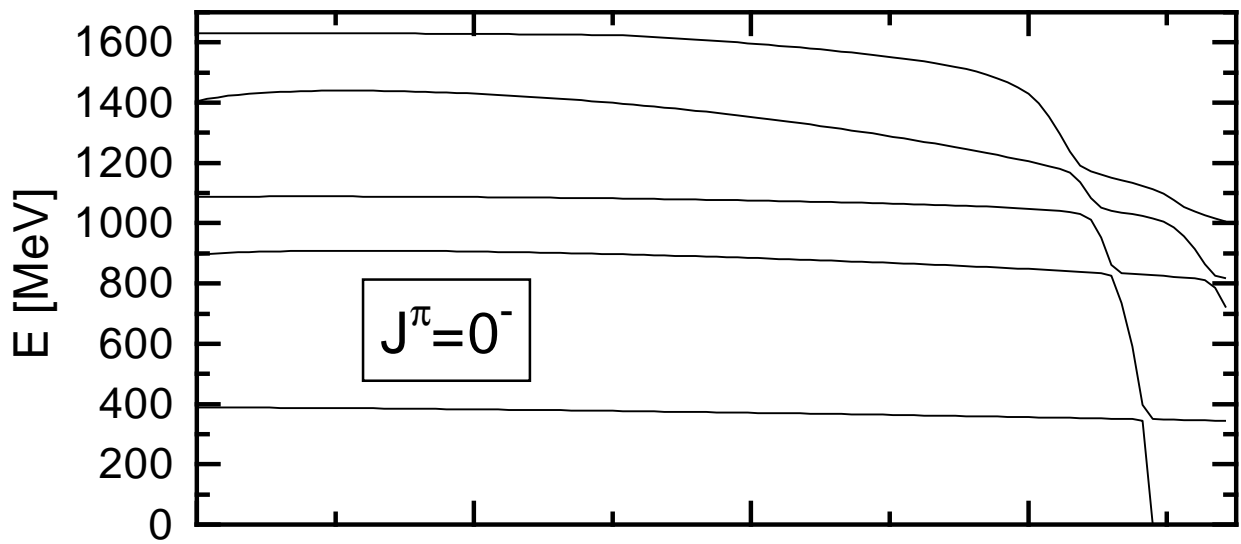
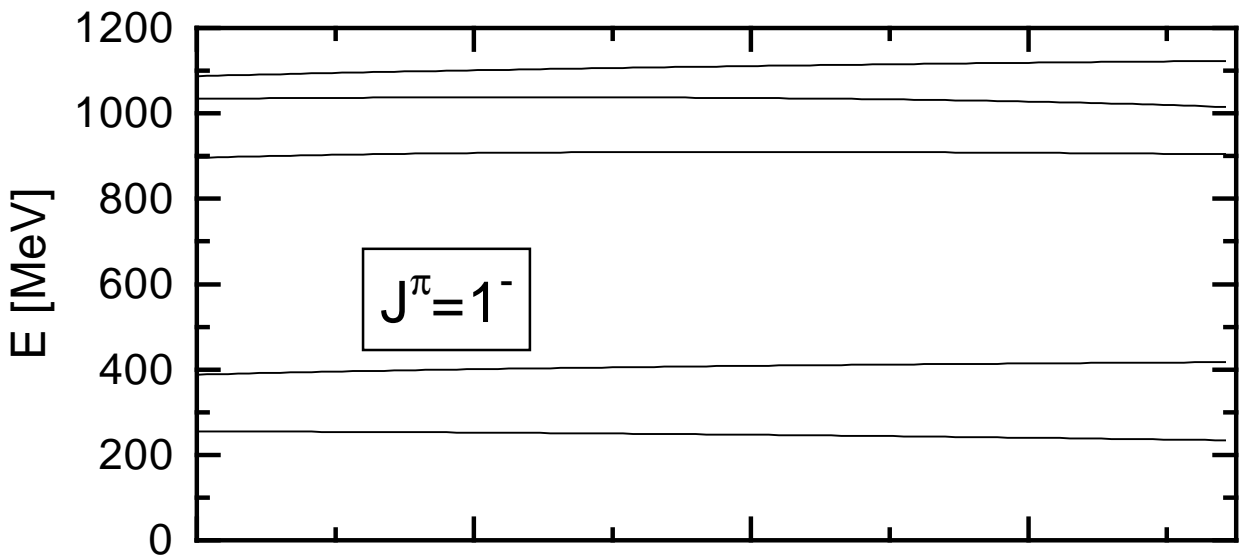
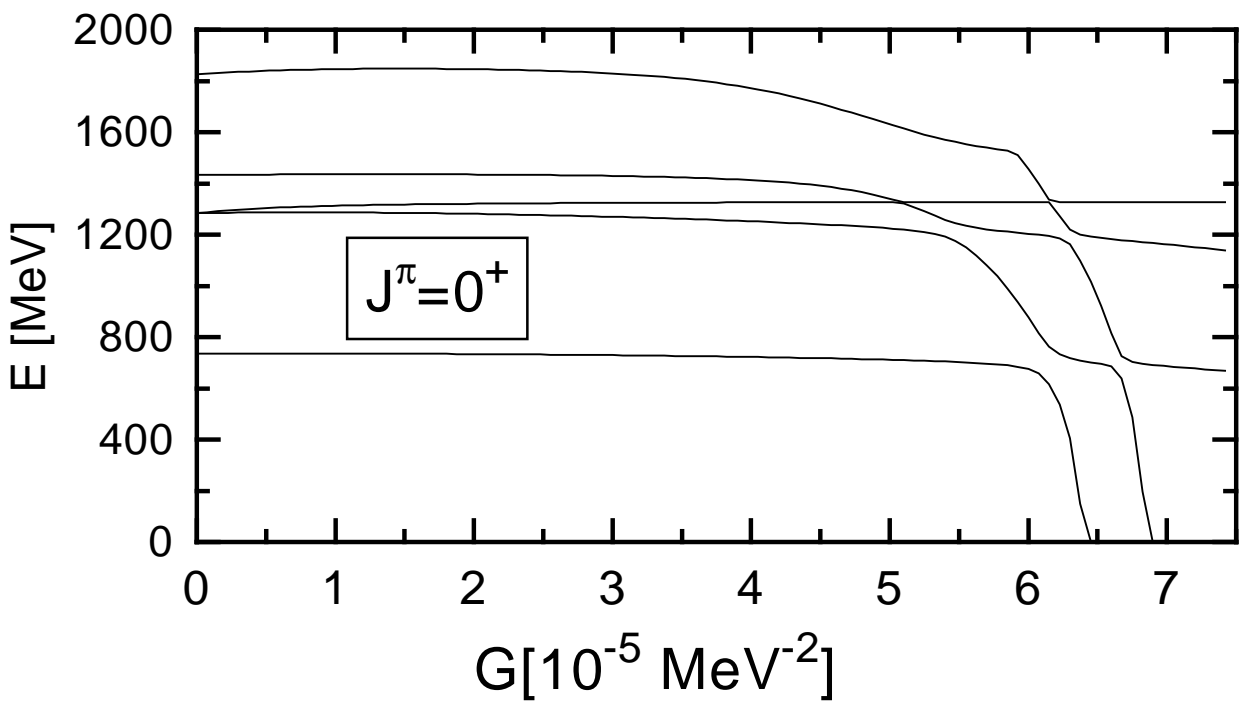
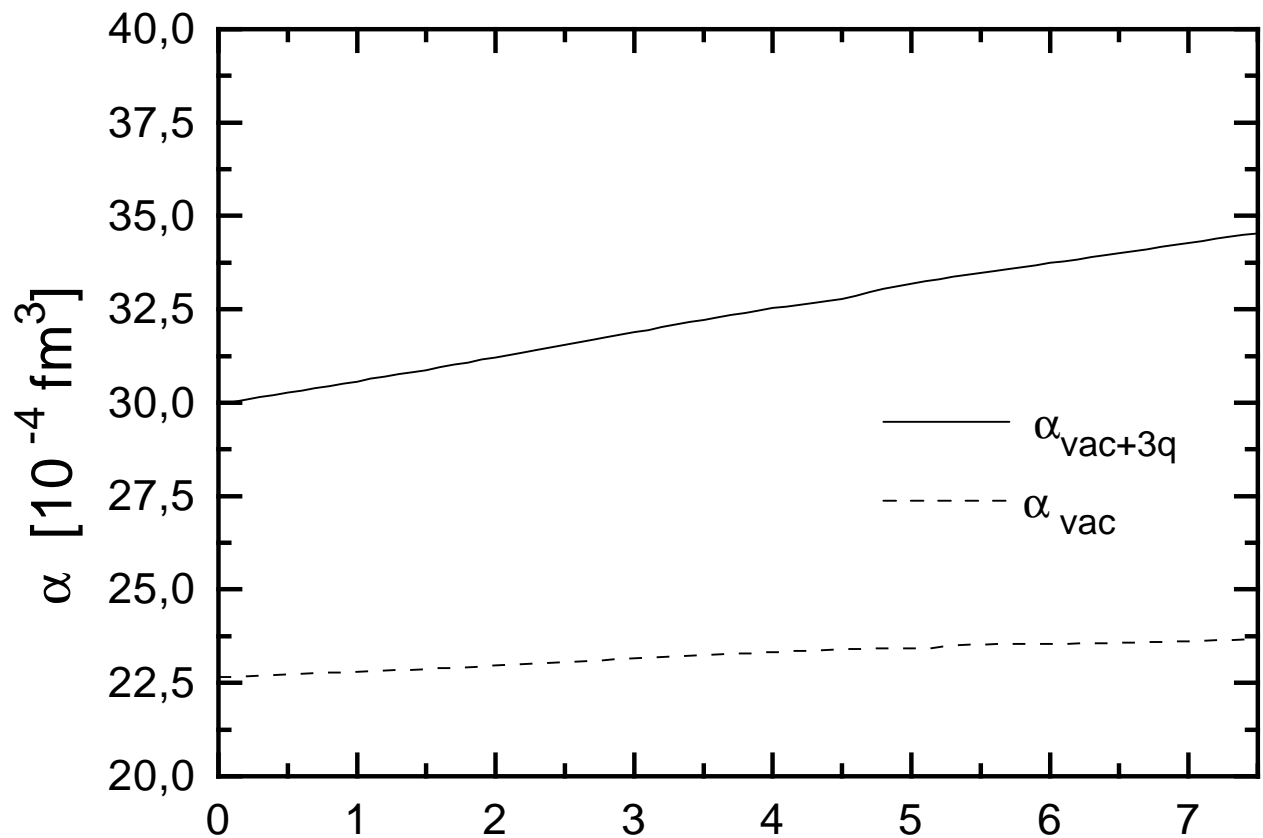
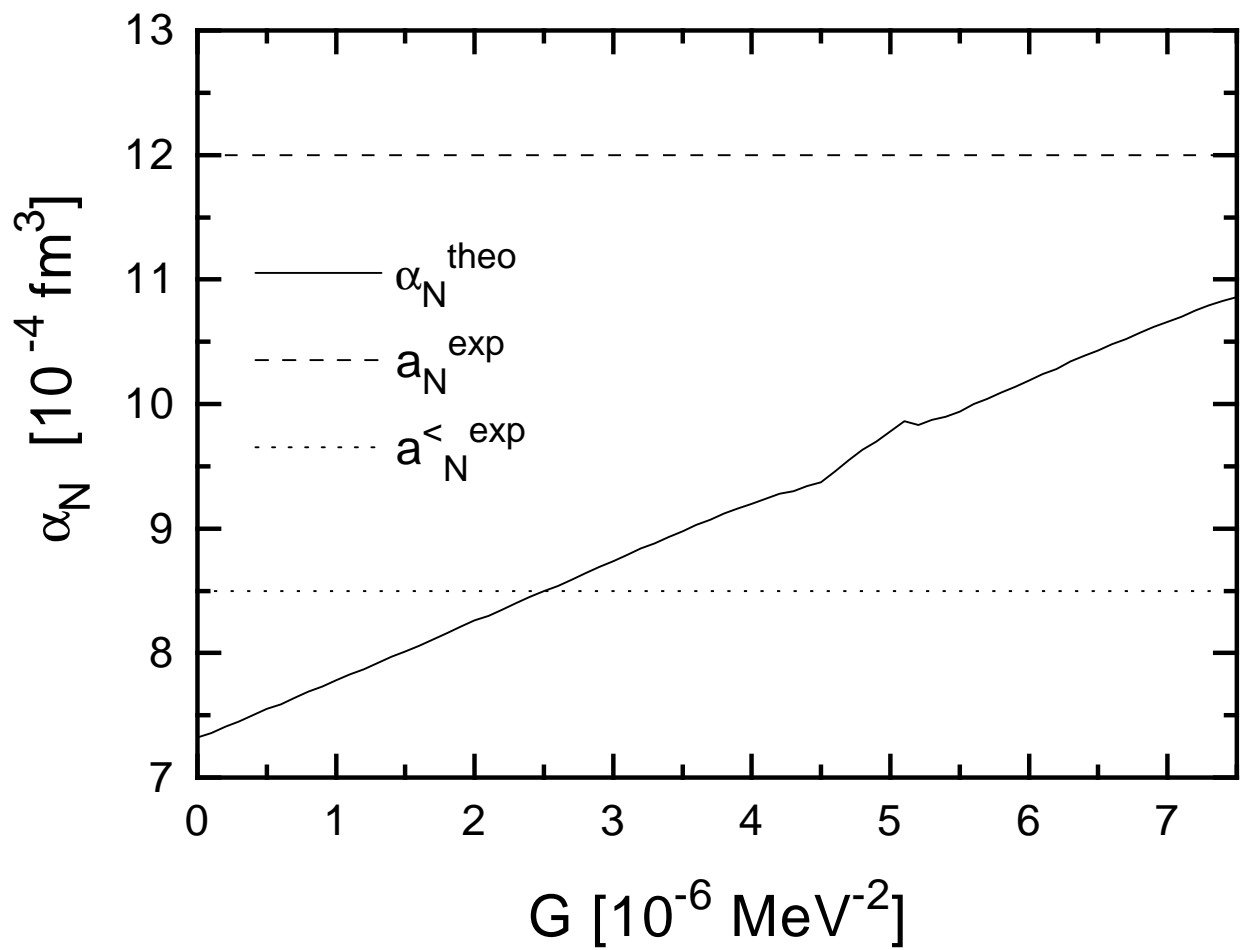
\end{document}